\documentclass[aps,twocolumn,showpacs,amsmath,amssymb,superscriptaddress]{revtex4-1}
\usepackage{graphicx}
\usepackage{dcolumn}
\usepackage{bm}
\usepackage{natbib}
\usepackage{color} 
\usepackage[normalem]{ulem}

\begin{document}

\title{Optical evidence of the chiral magnetic anomaly in Weyl semimetal TaAs}

\author{A. L. Levy}
\affiliation{Center for Nanophysics and Advanced Materials, Department of Physics, University of Maryland, College Park, Maryland 20742, USA}\affiliation{National Institute of Standards and Technology (NIST), Gaithersburg, Maryland 20899, USA}
\author{A. B. Sushkov}
\affiliation{Center for Nanophysics and Advanced Materials, Department of Physics, University of Maryland, College Park, Maryland 20742, USA}\author{Fengguang Liu}
\affiliation{Center for Nanophysics and Advanced Materials, Department of Physics, University of Maryland, College Park, Maryland 20742, USA}\affiliation{Department of Materials Science and Engineering, Anhui Agricultural University, Hefei 230036, China }
\author{Bing Shen}
\author{Ni Ni}
\affiliation{Department of Physics and Astronomy and California NanoSystems Institute, University of California, Los Angeles, CA 90095, USA}
\author{H. D. Drew}
\affiliation{Center for Nanophysics and Advanced Materials, Department of Physics, University of Maryland, College Park, Maryland 20742, USA}
\author{G. S. Jenkins}
\affiliation{Laboratory for the Physical Sciences, College Park, Maryland, 20740}
\date{\today}

\begin{abstract}
	Chiral pumping from optical electric fields oscillating at THz frequencies is observed in the Weyl material TaAs with electric and magnetic fields aligned along both the $a$- and $c$-axes. 
Free carrier spectral weight enhancement is measured directly for the first time, confirming theoretical expectations of chiral pumping. A departure from linear field-dependence of the Drude weight is observed at the highest fields in the quantum limit, providing evidence of field-dependent Fermi velocity of the chiral Landau level. Implications for the chiral magnetic effect in Weyl semimetals from the optical $f$-sum rule are discussed.
\end{abstract}

\maketitle
\section{Introduction}
Weyl fermions are massless, chiral particles that play an important role in electroweak interactions but have yet to be observed. Recent theoretical development in topological condensed matter systems predict emergent Weyl excitations described by the same underlying physics \cite{nielsen1983adler,XuScience2015,Wan2011,Xu2011,Xu2016}. These excitations in Weyl semimetals display unique physical phenomena including the existence of pairs of 3D Dirac cones with indicative spin textures near nodes, Fermi arc surface states \cite{Xu2011}, enhanced longitudinal magneto-conductivity,\cite{Son2013} and novel plasmonic \cite{HofmannDasSarma2015} and photovoltaic properties \cite{Ma2017Nature,osterhoudt2017colossal}.
              
Despite various experimental observations consistent with the predicted properties of the emergent Weyl states \cite{Xiong413,Huang2015,zhang2016signatures,Zhang2016PRB}, evidence of one of its key characteristics, the chiral pumping effects arising from the Adler-Bell-Jackiw chiral anomaly, is not certain. The chiral pumping effect is predicted to result in an enhanced free-carrier Drude spectral weight when the electric and magnetic fields are applied in parallel ($e\parallel B$) \cite{nielsen1983adler,HofmannDasSarma2015}. Bulk transport experiments report a negative longitudinal magneto-resistance \cite{Xiong413,Huang2015,zhang2016signatures,Zhang2016PRB}, qualitatively consistent with chiral pumping. However, effects from current jetting and off-diagonal components of the magneto-resistivity tensor are known to cause negative longitudinal magneto-resistance for systems with highly mobile charge carriers, leaving the observation of chiral pumping under intensive debate \cite{pippard1989magnetoresistance,CurrentJetting,Zhang2016PRB,Liang2018}. Furthermore, the Weyl scattering rate is expected to depend on magnetic field, obfuscating direct comparison between electronic transport data and predicted free-carrier Drude weight enhancements. This is particularly problematic in the quantum limit (QL), where the enhancement of the longitudinal magneto-resistance observed in electronic transport \cite{Huang2015,Zhang2016PRB} has been attributed to nontrivial changes in both the Drude weight and scattering rate under magnetic field \cite{Lu2015}.

All of these issues are addressed by performing the first broadband magneto-reflectance measurements that detect chiral pumping effects at terahertz (THz) frequencies. Current jetting effects are circumvented because optical measurements do not require contacts, opening the possibility of measurements along multiple crystal axes on the same crystal. Weyl Drude weight enhancement and scattering rate are independently characterized by extending transport measurements into the frequency domain. Direct observation of a Drude weight enhancement using light whose polarization is parallel to an applied field is especially significant since no optical signature is expected from trivial (non-Weyl) carriers \cite{palik1970infrared}.

Magneto-optical results are presented in the Weyl system TaAs in the 
$e\parallel B\parallel a$ and $e\parallel B\parallel c$ geometries, where $e$ is the polarization direction, $B$ is the statically applied magnetic field, and $a$ and $c$ are the crystal axes. A magnetic field-induced Drude weight enhancement, a hallmark signature of the chiral magnetic anomaly, is observed for the first time in both geometries. Section II discusses optical selection rules and expected changes in reflectance for an ideal Weyl semimetal in the $e\parallel B$ geometry. In Section III, the optical response at zero field is extracted from reflectance spectra for both $e\parallel a$ and $e \parallel c$. Section IV and V present data and analysis of reflectance spectra for $\parallel B\parallel a$ and $e\parallel B\parallel c$,respectively. Drude spectral weight enhancement is observed and discussed. Evidence of field-dependence of the Fermi velocity is also observed in both crystal orientations and naturally explained using the optical $f$-sum rule.
\section{Expected optical response from an ideal Weyl Semimetal}


\begin{figure}[htpb]
	\includegraphics[width=\columnwidth]{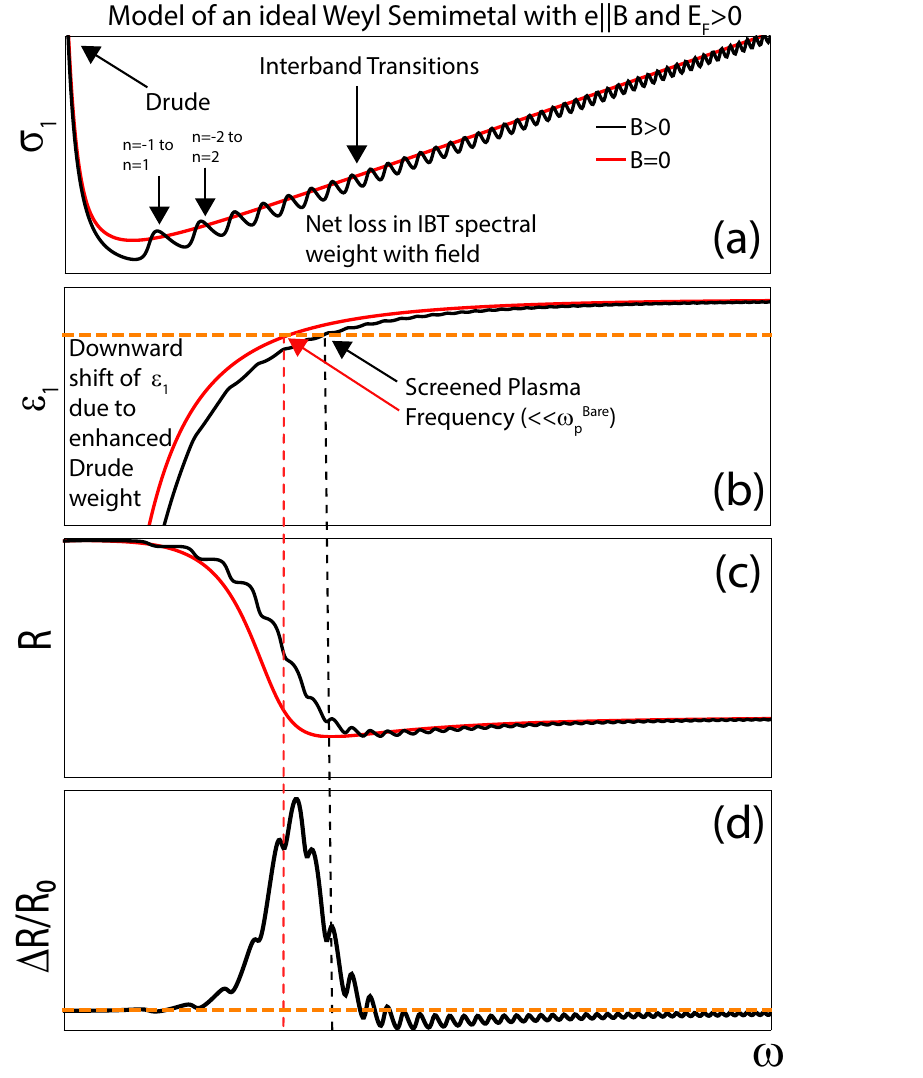}
	\caption{The expected optical response and reflectance expected for an ideal Weyl semimetal with the scattering rate much less than the lowest measured frequency are shown. (a) The real part of the optical conductivity ($\sigma_1$) as a function of frequency ($\omega$) is shown for an ideal Weyl semimetal at zero magnetic field (red) and finite magnetic field (black). The optical conductivity at finite field was obtained using Eqs. \ref{eq:sigma} and \ref{eq:Selection}. The oscillations in $\sigma_1$ at finite field are due to interband transitions from the $n^{\textnormal{th}}$ Landau level of the valence band to the $n^{\textnormal{th}}$ Landau level of the conduction band. (b) The real part of the dielectric response $\varepsilon_1(\omega)$ for the ideal Weyl semimetal at zero magnetic field and finite magnetic field are shown in red and black, respectively. A horizontal, dashed orange line demarcates $\varepsilon_1=0$. The frequency at which $\varepsilon_1=0$ is the screened plasma frequency, which is visually represented by the dashed red and black vertical lines for $B=0$ and $B>0$, respectively. This is obtained using a Kramers-Kronig transformation of $\Delta\sigma_1$. (c) Reflectance is calculated from the contributions to the optical response shown in (a) and (b) arising from Weyl interband transitions and free-carrier response for zero and finite field (red and black, respectively). The characteristic shift in the reflectance (screened) plasma edge gives a measure of the Drude weight enhancement. (d) The relative change in reflectance ($(R_B-R_0)/R_0$) is shown in black alongside a dashed orange line at $\Delta R/R_0=0$. }
	\label{fig:modelreflectance}
\end{figure}

Reflectance in the $e\parallel B$ Voigt geometry is rarely used to study material properties because expected signals are typically very weak compared to other geometries. Since this geometry is not typical, a review of the expected optical signals is presented for an idealized Weyl system.

This section consists of two parts. The first explains the selection rules in the $e\parallel B$ Voigt geometry. In addition to showing expected changes in the optical conductivity of interband transitions under magnetic field, these selection rules show why a cyclotron mode is not expected in this geometry. The change to the Drude weight due to Weyl pockets are then discussed in the context of the $f$-sum rule. 

The second part discusses the reflectance resulting from changes in the Weyl interband and Drude contributions to the optical response. Section \ref{Sec2} discusses the expected changes in conductance and reflectance with magnetic field in this geometry. Negligible optical signals produced by conventional carriers is the underlying reason that this geometry is not typically used to study material systems.

\subsection{Selection rules for $e\parallel B$}
The $e\parallel B$ Voigt geometry is rarely used since no change in reflectance or transmission is expected if oriented along the principal axis of a Fermi surface \cite{palik1970infrared}. For normal-incident light polarized along the principal axis of a crystal in the $e\parallel B$ geometry, the reflectance is given by:
\begin{equation}
R=\left|\frac{\sqrt{\varepsilon_{zz}(\omega)}-1}{\sqrt{\varepsilon_{zz}(\omega)}+1}\right|^2
\label{reflectance}
\end{equation}
where $R$ is the reflectance, $\epsilon(\omega)$ is the dielectric response, and $z$ is the direction of the electric field polarization. 

The selection rules for a Weyl semimetal in the $e\parallel B$ Voigt geometry with a magnetic field applied along the z-direction are discussed following the derivation reported in Refs. \cite{ashby2014PRB,shao2016magneto}. When the ideal Weyl semimetal (described in Eq. \ref{Hamiltonian} in Appendix A), is subjected to magnetic fields along the z-direction, the wave functions close to the Weyl point $k=\left(0,0,k_0\right)$ take the following form:

\begin{equation}
{\left|\psi_{n,b}\right>=
	\left|n\right>\otimes\left|	\begin{array}{cc}	
	0\\
	C_{\uparrow nb}\\
	\end{array}\right>
	+b\left|n-1\right>\otimes\left|	\begin{array}{cc}	
	C_{\downarrow nb}\\
	0\\
	\end{array}\right>
}
\label{S2}
\end{equation}
where $n\geq0$ is the Landau level (LL) index, $b=+(-)1$ denotes states in the conduction (valence) band, and the spin s along the magnetic field is denoted by $\uparrow$ or $\downarrow$. Coefficients $C_{snb}$ are given by \cite{shao2016magneto}:

\begin{equation}
{C_{\uparrow(\downarrow)nb}=\sqrt{\frac{1}{2}\left(1+(-)\frac{-\frac{\Delta}{2}+\frac{\Delta}{2 k_0^2}k_z^2}{b\sqrt{\frac{2n\hbar^2 v_x v_y}{l_B^2}+\left(-\frac{\Delta}{2}+\frac{\Delta}{2 k_0^2}k_z^2\right)^2}}\right)}}
\end{equation}
$\Delta$ is the energy gap at the saddle point between the Weyl points, which are separated in k-space by $k_0$. The field is taken along the line node separating the two Weyl points $k$-space for algebraic simplicity. The wavefunction of the chiral $n=0$ LL is given by $\left|\psi_0\right>=\left|0\right>\otimes\left|\begin{array}{cc} 0\\ 1\\ \end{array}\right>$.  The magneto-optical conductivity can be approximated through the Kubo formula:
\begin{widetext}
	\begin{equation}
	\label{eq:sigma}
	\sigma_{\alpha\beta}(\hbar\omega)=-\frac{i\hbar}{2\pi l_B^2} \sum_{n,n',b,b'} \int\frac{dk_z}{2\pi}\frac{f(E_{nb})-f(E_{n'b'})}{E_{nb}-E_{n'b'}}\frac{\left<\psi_{nb}\right|j_\alpha\left|\psi_{n'b'}\right>\left<\psi_{n'b'}\right|j_\beta\left|\psi_{nb}\right>}{\hbar\omega-(E_{n'b'}-E_{nb})+i\hbar\Gamma}
	\end{equation}
\end{widetext}
where $f(E)$ is the Fermi-Dirac distribution function, $\Gamma$ is the scattering rate, and $j_\alpha$ are the current density operators are given by:
\begin{equation}
j_\alpha= e \frac{\partial H}{\partial \Pi_\alpha}
\end{equation}

with $\Pi_\alpha=p_\alpha-e A_\alpha/c$. For the ideal case, this gives:\newpage

\begin{equation}
\begin{array}{ll} 
j_x =&e v_x \sigma_x \\
j_y =&e v_y \sigma_y \\
j_z =&e\frac{\hbar k_z}{m^*}\sigma_0+\frac{e\Delta}{\hbar k_0^2}k_z\sigma_z\\
\end{array}
\end{equation}
For $e\parallel B$, the current density matrix elements take the form: 
\begin{widetext}
	\begin{equation}
	\label{eq:Selection}
	\begin{array}{ll}
	\left<\psi_{n'b'}\right|j_z\left|\psi_{nb}\right>&=e\left(\frac{\hbar k_z}{m^*}+\frac{\Delta k_z}{\hbar k_0^2}\right)C_{\downarrow n'b'}^*C_{\downarrow nb}\delta_{n,n'}+e\left(\frac{\hbar k_z}{m^*}+\frac{\Delta k_z}{\hbar k_0^2}\right)b'b C_{\uparrow n'b'}C_{\uparrow nb}^*\delta_{n'n}\\
	&=e\frac{\Delta k_z}{\hbar k_0^2}\left(C_{\downarrow n'b'}^*C_{\downarrow nb}-b b'C_{\uparrow n'b'}C_{\uparrow nb}^*\right)\delta_{n,n'}+e\frac{\hbar k_z}{m^*}\left(C_{\downarrow n'b'}^*C_{\downarrow nb}+b b'C_{\uparrow n'b'}C_{\uparrow nb}^*\right)\delta_{n'n}\\
	& =e\left[\frac{\Delta k_z}{\hbar k_0^2}\sqrt{\frac{\frac{2n\hbar^2 v_x v_y}{l_B^2}}{{\frac{2n\hbar^2 v_x v_y}{l_B^2}+\left(-\frac{\Delta}{2}+\frac{\Delta}{2 k_0^2}k_z^2\right)^2}}}\delta_{b',-b}+\left(\frac{\hbar k_z}{m^*}-b\frac{\Delta k_z}{\hbar k_0^2}\right)\sqrt{\frac{\left(-\frac{\Delta}{2}+\frac{\Delta}{2 k_0^2}k_z^2\right)^2}{\frac{2n\hbar^2 v_x v_y}{l_B^2}+\left(-\frac{\Delta}{2}+\frac{\Delta}{2 k_0^2}k_z^2\right)^2}}\delta_{b',b}\right]\delta_{n,n'}\\
	\end{array}
	\end{equation}
\end{widetext}

These selection rules differ significantly from those for Weyl pockets in the $e\perp B$ geometry \cite{shao2016magneto}. The first term in the last line corresponds to interband transitions (IBTs) while the second is the Drude response. All terms are proportional to $\delta_{n,n'}$, so all IBTs must conserve LL index and there are no cyclotron (gapped $\delta_{b,b'}$) modes for $e\parallel B\parallel z$. This is also the case for ellipsoidal pockets with quadratic dispersion, as the strength of the cyclotron mode is proportional to $\vec{j}\times \vec{B}$ in the pocket. While the curvature of the Fermi surface of the hole pockets could lead to a small nonzero contribution, it is expected to be negligible for $e\parallel B\parallel a$ and $e\parallel B\parallel c$ owing to the small angle of curvature and small proportion of spectral weight involved in the off-diagonal response \cite{palik1970infrareda}. This point is discussed further in Section IV(B). 

Changes in the optical response from trivial carriers are also expected to be negligible \cite{palik1970infrareda}. This is especially true of the trivial hole carriers in TaAs, whose large Fermi surface area perpendicular to the direction of the applied fields ensures that they will be in the classical limit \cite{LeeBansil2015}. Changes in both the IBT and Drude contributions to the reflectance are therefore expected to be dominated by Weyl carriers.

\subsection{Change in reflection spectra with magnetic field for $e\parallel B$\label{Sec2}}


The change in the optical conductivity from interband transitions of the Weyl pockets is illustrated in Fig. \ref{fig:modelreflectance}(a), which shows the optical conductivity at zero field (red) and finite field (black). A finite Fermi level Pauli-blocks optical transitions in a Weyl pocket at $\omega=2 E_F/\hbar$. Fermi's golden rule applied to a 3D Dirac band results in a linearly-increasing real conductivity ($\sigma_1(\omega)$) with frequency. Landau levels form with the application of a magnetic field, and the spacing between them increases with field. The lowest Landau level transition in the $e\parallel B$ Voigt geometry is the $-1\rightarrow 1$ transition; no IBTs involve the chiral n=0 Landau level. The lost $\sigma_1$ spectral weight from the interband transitions as magnetic field increases is compensated by an equal increase in the Drude weight, thereby satisfying the $f$-sum rule. At small fields, the Weyl Drude weight is theoretically predicted to increase quadratically with field, but rolls over to a linear behavior at higher fields \cite{Nielsen1983,Son2013,Hofmann2016}.

Changes in the Drude weight are measurable through $\sigma_1$ or $\varepsilon_1$. If the optical probe measures frequencies comparable to the scattering rate, then the Drude conductivity can be spectrally resolved in $\sigma_1$. However, if this is not the case and the lowest probed frequency is much greater than the scattering rate, the strength (spectral weight) of the Drude response is measurable by $\varepsilon_1$. This approach has typically been used to optically characterize the free-carrier (Drude) condensate weight at zero frequency in superconductors.

In simple metallic systems with only a Drude conductivity contribution, the dielectric function $\varepsilon$ near the screened plasma frequency, where $\varepsilon$ crosses zero, shown for both zero and finite field in Fig. \ref{fig:modelreflectance}(b), results in a local minimum in the reflectivity preceded by a sharp rise in reflectivity that must approach 100$\%$ at low frequency. A hypothetical increase in the Drude weight causes the zero crossing to shift to higher frequency.

The Weyl reflectance derived from the expected Drude and IBT optical response contributions with and without applied field is shown in Fig. \ref{fig:modelreflectance}(c). The overall behavior is metallic. The screened plasma frequency blue-shifts with increasing field (Fig. \ref{fig:modelreflectance}(b)), thereby shifting the reflectance edge (Fig. \ref{fig:modelreflectance}(c)). The small deviation between the screened plasma frequency and the minima in reflectance is due to the presence of IBT contributions to $\sigma_1$ at the screened plasma frequency.

Fig. \ref{fig:modelreflectance}(d) shows the change in reflectance with applied field, $\Delta R/R=\frac{R(B)-R_0}{R_0}$. The peak corresponds to the steepest part fo the reflectance edge that blue-shifts with the screened plasma frequency as shown in Fig. \ref{fig:modelreflectance}(c.)

The oscillations in reflectance above the plasma edge arise from local maxima in the joint density of states associated with interband transitions. They are broadened by finite scattering and transition rates, decreasing their already small signature.



\section{Measured optical response in zero field}
\begin{figure*}[ht]
	\includegraphics[width=\textwidth]{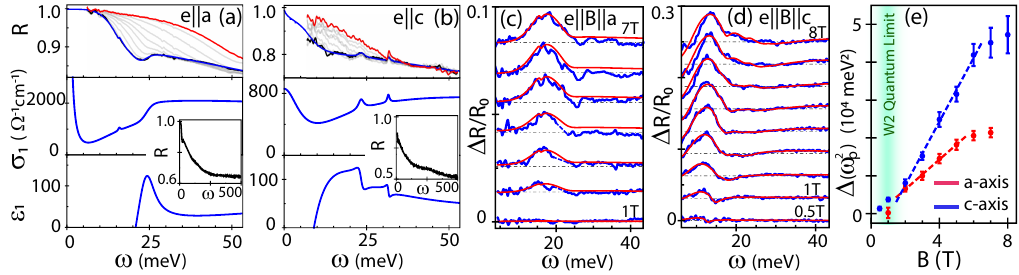}
	\caption{ (Color online).
		(a) The top panel reports reflectivity spectra R performed on a TaAs single crystal in the $ab$-plane at zero field in the $e\parallel a$ geometry. Multiple temperatures are reported that range from 300 K (red) to 10 K (black) with a superimposed fit (blue). The resulting low-temperature optical response functions are shown in the lower two panels. The inset shows the reflection taken over a broader spectral range. (b) Similarly, the zero field reflectivity data and optical response functions are reported for the $e \parallel c$ geometry in the $ac$-plane. (c) Magnetic field induced changes in reflectivity $\Delta R$, normalized by the zero field value $R_0$, are measured in the Voigt geometry on the as grown $ab$-facet with $e\parallel B\parallel a$ in the $ab$-plane at 10K (blue) and fit (red). A dominant peak appearing near the screened plasma frequency is indicative of the chiral pumping effect. Spurious noise caused by 60 Hz pick-up was removed from the 2, 3, and 4T data in a 2meV band centered at 21 meV, caused by 60Hz pick-up, was removed from the 2, 3, and 4T spectra and replaced with a linear interpolation (thin blue dashed line). (d) $\Delta R/R_0$ data is reported with $e\parallel B\parallel c$ on a polished $ac$-plane. (e) The enhanced Drude weight, reported as the change in the bare plasma frequency $\Delta(\omega_p^2)$, is derived from the fits in (c) and (d) and plotted as a function of magnetic field. The reported $a$-axis values are interpreted as lower bound values, which underestimate the enhanced Drude weight by no more than 20$\%$.
	}
	\label{fig:conductivity}
\end{figure*}

Magneto-optical measurements are performed using Fourier transform infrared (FTIR) spectroscopy. TaAs is a low-doped Weyl semimetal \cite{XuScience2015}. It is non-centrosymmetric with two types of Weyl chiral-conjugate pairs: four pairs of one type (W1) and eight pairs of another type (W2) \cite{XuScience2015, murakami2007phase,Wan2011}. The W1 nodes are 13 meV lower in energy than W2 nodes, leading to smaller relative chemical potentials above the node in the W2 pockets \cite{XuScience2015,yang2015weyl,Lv2015,Huang2015,LeeBansil2015,arnold2016}. The total free carrier response is a combination of these 24 Weyl pockets, and 8 identical trivial hole pockets. 

Single crystals of TaAs are grown by chemical vapor transport method as described in Ref. [10]. Optical measurements on a thick single crystal are reported on the as-grown $ab$-facet versus a mechanically polished $ac$-plane, which accommodates a 5 mm and 1.8 mm diameter aperture, respectively. FTIR spectroscopy measurements are performed at normal incidence in fields up to 8 T in the e$\parallel$B Voigt geometry, results that will be presented in Sections IV and V.

In the semimetal TaAs, field-induced enhancements arising from the chiral pumping effect contribute to $\varepsilon_1$ and shift the zero crossing to higher frequencies thereby shifting the reflectivity edge through the mechanism discussed in the previous section. In our work, this change in reflectivity in the vicinity of the plasma edge is used to sensitively detect chiral pumping effects as illustrated in Fig. \ref{fig:conductivity}(c). The total complex dielectric function in zero field is required to quantitatively extract the field-induced Drude weight enhancements from the shifting plasma edge. Zero-field reflectance measurements are performed to characterize the complex spectral dielectric function along the $a$- and $c$-axes. Zero-field and Faraday ($e \perp B$) geometry optical measurements also provide supplemental characterization of average band velocities, energy of the chemical potential above the Weyl nodes, and the crossover field range separating the classical and quantum limits.
 	
Reflectance spectra on the $ab$-plane are reported in the top panel of Fig. \ref{fig:conductivity}(a). The data are similar to previously published data \cite{Xu2016}. The measured reflectivity spectra are fit over a sufficiently large spectral range using a sum of dielectric Lorentzian oscillators that obey Kramers-Kronig relations \cite{Kuzmenko2005}. Note that this method gives a close approximation of the total optical response \cite{Kuzmenko2005}; the fitting parameters used are not all physically meaningful on their own. 

\subsection{Optical Response for $e\parallel a$ and $B$=0}
The model reflectance of a semi-infinite slab is $R~=~\left|\frac{1-\sqrt{\varepsilon(\omega)}}{1+\sqrt{\varepsilon(\omega)}}\right|^2$,  where the complex dielectric function is given by:
\begin{equation}
\label{0FieldDielectricEquation}
\varepsilon(\omega)=\varepsilon_{\infty}+   \sum_j\frac{\omega_{sj}^2}{\omega_{0j}^2-\omega^2-i\omega\gamma_j}+\varepsilon_G(\omega)
\end{equation}
The complex optical conductivity is related to the dielectric function by $\sigma= {\omega\varepsilon}/{4\pi i}$. The first term $\varepsilon_\infty$  is a constant arising from cumulative interband contributions at frequencies greater than $500$ meV. The second term includes a zero-frequency (Drude term) oscillator, a phonon oscillator centered 15.9 meV and a very broad additional oscillator used to fit the background of interband transitions. 
An interband transition (IBT) feature in the reflectivity data near 25 meV requires a third term $\varepsilon_G(\omega)$ that has a sharp absorptive onset followed by a $\sim 1/\omega^3$ frequency roll-off. A model of $\varepsilon_G(\omega)$ with the requisite behavior is given by
 $\varepsilon_{1G}(\omega)=1+8\int_0^{\infty}\frac{\sigma_1(\omega^\prime)d\omega^\prime}{\omega^{\prime2}-\omega^2}$,
~$\sigma_1(\omega)=\Gamma(\omega)\sigma_{1D}(\omega)$~,
~$\Gamma(\omega)=0.5[1+\textnormal{tanh}(\frac{\omega-2\Delta}{2\delta_\Delta})]$, and
~$\sigma_{1D}(\omega)=\frac{\omega_p^2\gamma}{4\pi(\omega^2+\gamma^2)}$. All terms are proper response functions that are Kramers-Kronig constrained, so the resulting fitted (total) $\varepsilon(\omega)$ is directly derived from the reflection data independent of the underlying model.


The fitted response functions, $\sigma_1(\omega)$ and $\varepsilon_1 (\omega)$ (parameters reported in Appendix C) are shown in the middle and bottom panels of Fig. \ref{fig:conductivity}(a), respectively. A linearly increasing $\sigma_1(\omega)$ is expected from the IBTs of an ideal 3D-Dirac cone above the transition onset frequency. Linear behavior is observed between 6 and 25 meV, suggesting the existence of an IBT onset energy below 6 meV. In an ideal Dirac cone, the $\omega_{onset}$ of an IBT is $\omega^{ideal}_{onset} = 2\left|E_F\right|/ \hbar$, where $E_F$ is the difference in energy between the chemical potential and the node. In asymmetric systems such as ours: if $E_F>0$, $\omega_{onset}$ is shifted below (above) $2E_F/\hbar$ if the Fermi velocity of the conduction band is larger (smaller) than that of the valence band, though must always be greater than $E_F/\hbar$. 
Since no onset freqeuncy is observed in the data down to 6 meV, the Fermi level is necessarily below this energy. Considering that the W2 pocket valence band-conduction band asymmetry is not severe \cite{LeeBansil2015,arnold2016}, the estimated Fermi level is less than $\sim 3$ meV above the node. This is consistent with reported Fermi levels in W2, whereas that of W1 is much larger \cite{LeeBansil2015,arnold2016}. A change in slope of $\sigma_1(\omega)$ is observed near 25 meV, which arises from the onset of IBTs in W1. Incorporating estimates of the degree of anisotropy in W1 from band structure calculations \cite{LeeBansil2015,arnold2016}, an onset frequency of 25 meV translates into a W1 Weyl node position that is $\sim$15 meV below the Fermi level.
 Therefore, considering that W1 is about 13 meV lower in energy than W2, the W1 and W2 nodes of our sample is 15 meV and $\sim 2$ meV below the Fermi level, respectively, agreeing with other experimental results \cite{XuScience2015,yang2015weyl,Lv2015,arnold2016}.  Above 25 meV, $\sigma_1$ remains unchanged with increasing frequency, suggesting a strong frequency roll-off of the $\sigma_1$ contributions from the Weyl states. This is attributed to the band curvature effects since intersecting pairs of Weyl states merge and form a (Lifshitz) gap, which is predicted to be $\sim 50$ meV for W1 and $\sim 80$ meV for W2. IBTs near the saddle point between the Weyl points are expected to have reduced optical transition matrix elements, causing the observed frequency roll-off of the conductivity \cite{Jenkins2016}.

\subsection{Optical Response for $e\parallel c$ and $B$=0}
Reflectance spectra on a polished $ac$-plane are reported in the top panel of Fig. \ref{fig:conductivity}(b). The $e\parallel c$ spectra are fit with a dielectric function similar to Eq. \ref{0FieldDielectricEquation}: $\varepsilon(\omega)=\varepsilon_{\infty}+\sum_j \frac{\omega_sj^2}{\omega_0j^2-\omega^2-i\omega\gamma_j}$, which includes a Drude term, a broad oscillator centered at 42.5 meV, and two phonon terms at 31.6 meV and 23.3 meV.\cite{RamanTaAs}  The fitted response functions $\sigma_1$ and $\varepsilon_1$ (parameters in Appendix C) are reported in the middle and bottom panels of Fig. \ref{fig:conductivity}(b), respectively.	

Figure \ref{fig:conductivity}(b) reports a $e\parallel c$ optical response that exhibits a weaker metallic behavior than for $e\parallel a$. The reduced free-carrier Drude weight lowers the screened plasma frequency and the associated reflectivity plasma edge inflection point from 20 meV ($a$-axis) to 10 meV ($c$-axis), and no W1 IBT feature is observed at 25 meV. These differences in optical response are attributed to hole and W1 pocket anisotropy and higher scattering rate from surface preparation. The $c$-axis Fermi velocity of the trivial holes and W1 carriers are both expected to be an order of magnitude smaller than that along the $a$-axis \cite{XuScience2015,yang2015weyl,Lv2015,arnold2016,LeeBansil2015}. Since the semiclassical Dirac Drude spectral weight is proportional to $n v_F/k_F$, the small $c$-axis Fermi velocity markedly reduces the spectral weight, suppressing the W1 IBT peak. In contrast, W2 pockets are much more isotropic, leading to a larger relative contribution to the total optical response for $e\parallel c$ \cite{XuScience2015,yang2015weyl,Lv2015,arnold2016,LeeBansil2015}.
Also, the Drude scattering rate for measurements on the same polished $ac$-plane, but with $e\parallel a$, is much larger than in the as-grown $ab$-plane, which is characteristic of blemished surface layers.

\section{$\lowercase{e}\parallel B\parallel a$ Voigt Geometry}

Normal-incident magneto-reflectance spectra reported in Figs. \ref{fig:conductivity}(c) and \ref{fig:conductivity}(d) are in the $e\parallel B \parallel a$ and $e \parallel B \parallel c$ geometry, respectively.
The B-field dependent reflectance spectra R(B) are normalized to the zero-field data $R_0$ and reported as $\Delta R(B)/R_0=R(B)/R_0-1$. Note that these results show qualitative agreement with the simulations for an ideal Weyl semimetal shown in Fig. \ref{fig:modelreflectance}(b). The dominant feature is an increasing positive peak with magnetic field, near the $R_0$ reflectance edge, indicative of a blue-shifting plasma frequency caused by an enhanced free carrier Drude weight. 

\subsection{Results for $e\parallel B\parallel a$}
\begin{figure}[tbh]
	\includegraphics[width=\columnwidth]{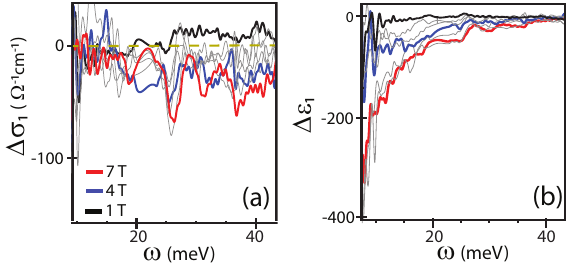}
	\caption{Shown in (a) and (b), respectively are $\Delta \sigma_1$ and $\Delta \varepsilon_1$ for $e\parallel B\parallel a$ in the measured spectral range obtained using $\delta R/R$ and its KK-transformation and the zero-field dielectric function. Drude enhancement manifests only $\Delta\varepsilon_1 $. $\Delta\sigma_1$ is dominated by changes to the optical response of the Drude and IBTs. A dark yellow dashed line is included in (b) to show that $\Delta\sigma_1<0$ in the measured spectral range, consistent with loss of IBT spectral weight. The change in optical response is dominated by $\Delta \varepsilon_1$.
	}
	\label{KK}
\end{figure}

Fig. \ref{KK} shows the change in optical response in the measured spectral range obtained from the relative reflectance spectra and their Kramers-Kronig transformations for $e\parallel B \parallel a$ \cite{KKmethod}. More information about this method and its justification is available in Appendix D and in Ref. [\citenum{KKmethod}]. 

The uncertainty in the change in optical response generated by the Kramers-Kronig method is~negligible as the change in relative reflectance $\Delta R/R_0$ is extremely small outside of the measured spectral range. The changes in R at low frequency are necessarily small since, for a metallic response, R is nearly 1 as shown in Fig. \ref{fig:conductivity}(a) as the response is dominated by the Drude. Changes in the optical response at low frequencies will therefore have a negligible effect on R; R will remain very close to 1 as long as TaAs remains metallic.  Fig. \ref{fig:modelreflectance} illustrates that in the ideal case, changes in the interband transition due to the application of magnetic field and the formation of Landau levels have a negligible effect on the optical response at high frequency leading to negligible changes to the reflectance. Therefore, at low and high frequencies outside the measure spectral range, $\Delta R$ tends towards 0.

The frequency-dependence of the large negative $\Delta\varepsilon_1$ and the small negative $\Delta\sigma_1$ support the interpretation that the change in optical response is dominated by Drude weight enhancement. The increasingly negative $\varepsilon_1$ at low frequency with field is indicative of a growing Drude weight. An increasingly negative $\sigma_1$ at high frequency indicates loss of interband spectral weight. Therefore, these two graphs show a transfer of spectral weight from high to low frequency. The increase of the Drude spectral weight at very low frequency is not directly observable since the scattering rate is small compared to the lowest measured frequency, as depicted in the ideal case in Fig. \ref{fig:modelreflectance}(a). This interpretation is consistent with the fact that $\Delta R(B)/R_0$ for $e\parallel B \parallel a$ is fit well by increasing the Drude weight of the zero-field optical response.

The $a$-axis data (red) in Figure \ref{fig:conductivity}(c) are fit (blue) using a single free parameter, an enhanced Drude weight $\Delta (\omega_p^2)/4\pi$ added to the zero-field Drude weight. The other zero-field model dielectric terms (whose parameters remain static) are also included in the $\Delta R(B)/R_0$ fit model. The monotonically increasing Drude weight with field is reported in Fig. \ref{fig:conductivity}(e). The increasing Drude weight is accompanied by increasing loss of IBT with field expected by the $f$-sum rule. At higher fields, the Drude weight increases more slowly with field, as shown in Figs. \ref{fig:conductivity}(e) and \ref{KK}(b). Drude weight enhancement for $e\parallel B\parallel a$ is observed through $\Delta\varepsilon_1$ and not $\Delta\sigma_1$ in the measured spectral range due to the small scattering rate.

The reflectance spectrum is more sensitive to $\Delta \varepsilon_1$ than $\Delta\sigma_1$ in the measured spectral range as it includes the plasma edge, where the reflectance is most sensitive to changes in $\varepsilon_1$.

\subsection{Discussion}

Before discussing the implications of these results on the Weyl pockets in TaAs, it is helpful to understand why this signal arises from effects related to the chiral magnetic anomaly and not from trivial hole pockets or cyclotron modes.

Selection rules in this geometry differ significantly from those of the Faraday and $e\perp B$ Voigt geometries, as discussed in Section II. Because the principal axes of the Fermi surfaces are oriented along the crystalline axes, cyclotron modes are not expected in these geometries.

The trivial hole pockets in TaAs are very anisotropic with curved Fermi surfaces that deviate from an ellipsoidal approximation whose principal axes are aligned along the crystal axes. Near the end points of the pockets where the deviations are largest, the curvature could, in principle create current that is not parallel to the driving electric field. However, estimates show that the number of carriers associated with these small areas compared with the volume of the pocket is negligible. The small curvatures of the Fermi surface, combined with the small spectral weight involved in such a process do not produce a significant effect. Experimentally, the observation of the blue-shift in plasma edge frequency with field along multiple crystal axes provides further evidence that Fermi surface curvature effects from trivial carriers are not responsible. 

Magnetic field dependence of the reflectance int he $e\parallel B\parallel a$ and $e\parallel B \parallel c$ Voigt geometries is expected from Weyl state carriers. The Drude weight and interband transitions have expected effects as discussed in Section II, Appendix A, and prior theoretical work \cite{nielsen1983adler,HofmannDasSarma2015,Spivak2016}.These effects are expected to be observable in the measured spectral range. Trivial (hole) carriers remain in the semiclassical limit at all fields applied, and interband transitions are not observed in zero field measurements below $500$ meV. Any measurable changes in the free carrier or interband transition spectral weight cannot be attributable to the trivial hole pockets in either Voigt geometry, especially increases in spectral weight.    


The W2 carriers are responsible for the peak feature in Fig. \ref{fig:conductivity}(c). The W1 carreires remain in the classical limit below 8T, a result derived from the relatively large chemical potential (as measured in zero-field) combined with Fermi velocity estimates. The QL for W2 carriers is expected to occur at $\sim$2T for our sample. Changes in the optical response for the spectral range $\omega>>\gamma$ from Weyl carriers in the semiclassical regime are minor compared to the QL,\cite{Son2013} where the Drude enhancement is independent of Fermi energy. Therefore, changes in the optical response are expected to be much stronger in W2 than W1 pockets \cite{nielsen1983adler,HofmannSarma2014,HofmannDasSarma2015,Spivak2016}. 

In the frequency range $\omega>>\gamma$, changes in the scattering rate have a negligible effect on the optical response. Changes to the optical response in the measured spectral range are dominated by changes in the Drude weight.

Figure \ref{fig:conductivity}(e) summarizes, in red, the magnetic field dependence in the $e\parallel B\parallel a$ geometry of the Drude weight enhancement, equal to $\omega_P^2/4\pi$ where $\omega_p$ is the plasma frequency. The Drude weight shows a linear field-dependence between 1 and 5 T, consistent with theoretical predictions of chiral pumping from a Weyl pocket in the QL \cite{nielsen1983adler,HofmannDasSarma2015,Spivak2016}:
\begin{equation}
\omega_p^2=v_F e^2/\hbar \pi l_B^2
\label{eq:ChiralDrude}
\end{equation}
where $l_B=\sqrt{\frac{\hbar c}{eB}}$. These results are consistent with the chiral pumping arising entirely from the W2 pockets that are in the QL at $2$T. The fitted slope for the $c$-axis data, 8400 meV$^2/$T, gives an average Fermi velocity of $1.2~\pm~0.15~\times~10^{-3} c$. A lower-bound average $a$-axis velocity of $0.58\pm0.15\times10^{-3} c$ is obtained from a fitted slope of 4000~meV$^2/$T in the $a$-axis data. This is close to the Fermi velocity expected from band structure calculations \cite{LeeBansil2015}. 

Above 5T, the Drude weight increases more slowly with magnetic field, becoming sublinear. The Drude weight in the quantum limit involves only the chiral n=0 Landau level. The slowing of the Drude enhancement, which indicates field-dependence in the n=0 Landau level Fermi velocity, is quantitatively related by the $f$-sum rule to the decreased rate of optical spectral weight lost from the interband transitions that do not involve the n=0 Landau levle. Realization of this relationship between the chiral free carrier response and the interband Landau level transitions provides new insight to both. 

The optical $f$-sum rule states that the spectral weight integrated over all frequencies remains constant provided the number of carriers does not change. In an ideal Weyl pocket in the $e\parallel B$ geometry, IBTs preserve Landau level index and do not involve the $n=0$ Landau level \cite{ashby2014PRB,shao2016magneto}. As derived in Appendix A, the resulting loss in IBT spectral weight is equal to the Drude weight enhancement that is exactly equivalent to the quantity in Eq. \ref{eq:ChiralDrude} found by other theoretical methods. 

In theoretical work that considers ideal Weyl states, the Weyl bands disperse linearly with frequency. In real material systems like TaAs, the intersection of the Weyl cones must give rise to a nonlinear band dispersion, resulting in deviations from the idealized behavior. For instance, the linear Weyl conductivity with frequency in the zero-field data presented in Fig. \ref{fig:conductivity}(a) and (b) is shown to roll over at high frequencies \cite{Jenkins2016}. As the IBT onset frequency (the n=-1 to n=+1 Landau level transition) increases with field and approaches the spectral range  where this roll over occurs, the loss of IBT spectral weight slows.  
By the $f$-sum rule, this decreased rate of IBT spectral weight loss translates into a slowed increase in Drude weight with magnetic field. From Eq. \ref{eq:ChiralDrude}, this decreased rate of Drude weight enhancement indicates a diminished $n=0$ Landau level Fermi velocity parallel to the field. 

The monotonic departure of the W1 and W2 contributions to $\sigma_1$ from linear dispersion in zero field measurements (most easily seen in Fig. \ref{fig:conductivity}(a)) indicates a departure from linear band dispersion. As the lowest allowed IBT transition energy approaches the frequency range in which $\sigma_1$ no longer increases linearly with frequency, a lower Fermi velocity is expected. \begin{flushright}
	
\end{flushright}This prediction is consistent with the increased longitudinal magnetoresistance observed at high fields in transport \cite{Xiong413,Huang2015,zhang2016signatures,Zhang2016PRB}.

While models that incorporate the Lifshitz saddle point between Weyl nodes predict a decrease in the Fermi velocity of the lowest Landau level at high fields \cite{Lu2015}, the presented measurements are the first to observed evidence of field-dependent Fermi velocity in the n=0 Landau level.

\begin{figure}[t]
\includegraphics[width=\columnwidth]{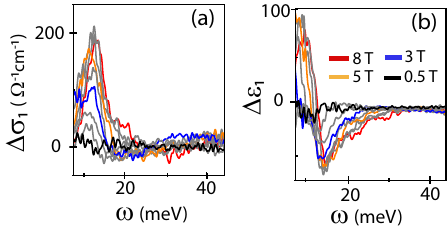}
\caption{Panels (a) and (b) report the changes in the optical response functions for $e\parallel B\parallel c$ derived from the $\Delta R/R_0$ data in Fig. 2(d) using a Kramers-Kronig transformation that incorporates the zero-field dielectric function.}
	\label{fig:KKCaxis}
\end{figure}


\section{$\lowercase{e}\parallel B\parallel \lowercase{c}$ Voigt Geometry}
\subsection{Sensitivity of $\lowercase{e}\parallel\lowercase{c}$ Optical Response}
Chiral pumping effects are optically observed in TaAs for $e\parallel B\parallel c$ on a polished $ac$-facet. Spectral evidence of an increasing transfer of spectral weight with field from Weyl interband transitions to the Drude contribution is presented. Results will be compared with the $e\parallel B\parallel a$ taking into account the highly anisotropic Fermi pockets. The polished $ac$-facet shows a much larger free-carrier scattering rate than the as-grown $ab$-facet allowing spectral resolution of the Drude response. The overall larger relative changes in optical response of the Drude as well as the IBT contributions allows models to separate the two contributions with minimal assumptions.

This larger relative changes in the reflectance for $e\parallel c$ with magnetic field is due to Fermi surface anisotropies. The trivial hole and W1 pockets are highly anisotropic, while W2 is much more isotropic. Both the hole and W1 pockets have greatly reduced Fermi velocities for $e\parallel c$ compared with $e\parallel a$. Both the Drude and IBT contributions to the total optical response are diminished. The relative contribution arising from the W2 pockets are therefore much larger.

Measurements of $\Delta R/R_0$ are therefore more sensitive to changes in W2. Since the QL of the W2 pocket occurs for B $\sim$ 2T independent of geometry, chiral pumping effects are expected to be dominated by W2. The $e\parallel B\parallel c$ spectra not only confirm the presence of the Drude enhancement from chiral pumping observed for $e\parallel B\parallel a$; they provide insights unavailable in the $e\parallel B\parallel a$ geometry.

\subsection{Results}
  
  \begin{figure}[t]
  	\includegraphics[width=\columnwidth]{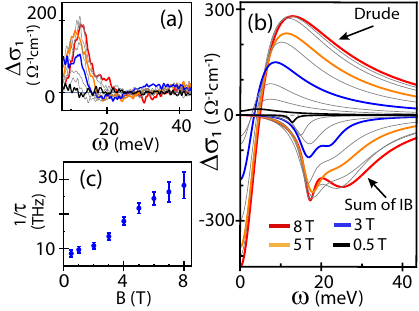}
  	\caption{Shown are the magnetic-field induced conductivity changes $\Delta\sigma_1$ associated with W2 Weyl pockets for the $\Delta R/R_0$ data in the $e\parallel B \parallel c$ geometry using (a) the KK-method, and (b) a physical model consisting of a replaced Drude term and two negative Lorentzians that represent the loss of spectral weight of the W2 Weyl IBTs. This model separates the Weyl Drude enhancement from contributions arising from the IBTs of W2 pockets. The sum of Drude and IBT terms (see Appendix C) gives similar results to the Kramers-Kronig analysis. (c) The magnetic field dependent inverse transport lifetime $1/\tau$ is reported.
  	}
  	\label{fig:fitting}
  \end{figure}
The enhanced $\Delta R/R_0$ signal for $e\parallel B \parallel c$ in Fig. \ref{fig:conductivity}(d) is much larger than for $e\parallel B \parallel a$ in Fig. \ref{fig:conductivity}(c) and the frequency and sign of the peak is consistent with a blue-shifting reflectance edge with field caused by a Drude weight enhancement of W2. Fig. \ref{fig:KKCaxis} shows the change in optical response in the measured spectral range obtained using a direct Kramers-Kronig analysis of the $\Delta R/R_0$ data. As Fig. \ref{fig:KKCaxis}(a) shows, $\Delta \sigma_1$ is larger in this geometry than in $e\parallel B\parallel a$ in the measured spectral range. 

Fig. \ref{fig:KKCaxis} shows that increasing spectral weight is transferred from high to low frequency with increasing field. The large negative increase in $\varepsilon_1$ is indicative of a growing Drude weight, and the broad suppression of $\sigma_1$ at high frequency is indicative of lost IBT spectral weight at high frequency. The very low frequency downturn of $\Delta\sigma_1$ and the positively growing $\Delta\varepsilon_1$ are indicative of scattering rate changes in the Drude as discussed in the next section. The larger signal-to-noise ratio of the $\Delta R/R_0$ $c$-axis data, reduced conductivity contribution from the trivial hole and W1 carriers, and serendipitously larger scattering rate all increase the reflectivity signals arising from magnetic field-induced changes in the W2 pocket. 

\subsubsection{Fitting with a Physical Model}

Since the reflectivity is much more sensitive to changes in W2 interband transitions in the $e\parallel B \parallel c$ geometry, the lost spectral weight from these IBTs must be taken into account. The change in optical response cannot be accurately modeled by a change in the Drude response alone.

The change in optical response is fit using a physical model with minimal degrees of freedom (the red fits in Fig. \ref{fig:conductivity}(d)), incorporating the zero-field dielectric function with two modifications: the Drude term is replaced with a new Drude term with two free fit parameters ($\omega_p^2,\gamma_0$), and two negative Lorentzians each with three free parameters ($\omega_{0j},\omega_{sj}^2,\gamma_{j}$). The sum of these two negative Lorentzians physically represent the loss of W2 IBT spectral weight expected for Weyl semimetals.



The negative $\Delta\sigma_1$ contributions at $\omega\lesssim$6 meV from the Drude term shown in Fig. \ref{fig:fitting}(b) are due to the low frequency negative tails with steep slopes in the $\Delta R/R_0$ spectra shown in Fig. \ref{fig:conductivity}(d). The small negative $\Delta R$ at low frequencies is an indication that the Drude scattering rate increases with spectral weight. The negative $\Delta\sigma_1$ contributions arising from the increasing scattering rate of the Drude term is not only necessary to fit the distinctive low frequency $\Delta R/R_0$ spectra, but is required to balance the net change in spectral weight to satisfy the $f$-sum rule, thereby validating the physical model.

The increase in scattering rate with field, reported in Fig. \ref{fig:fitting}(c) as the inverse transport lifetime $1/\tau$, is due to two effects. Blemishes from the mechanical polish extend over a characteristic depth. As the Drude conductivity increases with field, the penetration depth decreases, emphasizing the higher scattered region closer to the surface. The polished layer introduces scattering centers that break translational invariance, resulting in strong intervalley and interpocket scattering. These impurity-like scatterers can induce large wavevector changes, and therefore reach all parts of the Brillouin zone. The density of states at the Fermi level of W2 pockets in the QL scales as $1/l_B^2\sim B$, so scattering also tends to increase with field.\cite{Lu2015} This scattering rate field-dependence is consistent with observations in transport that the longitudinal magnetoresistance increases with field in the quantum limit{\tiny } \cite{Xiong413,Huang2015,zhang2016signatures,Zhang2016PRB}.

\subsection{Discussion}

The fact that the blue-shift of the plasma edge is also observed for $e\parallel B\parallel c$ provides further evidence that the blueshift in the $e\parallel B\parallel a$ geometry arises from enhancement of the Drude weight as predicted for Weyl semimetals. 

Figure 2(e) summarizes the enhancement of $\omega_p^2$ with field obtained from the physical model for $e\parallel B \parallel c$ in blue. These results are consistent with the chiral pumping arising entirely from the W2 pockets that are in the QL at $2$T. The fitted slope for the $c$-axis data, 8400 meV$^2/$T, gives an average Fermi velocity of $1.2~\pm~0.15~\times~10^{-3} c$. As with the $a$-axis resulst, the Fermi velocity is close to the expected value from band structure calculations \cite{LeeBansil2015}. The Fermi velocity along the $c$-axis is found to be significantly larger than the average velocity in the $ab-$plane. This is consistent with the fact that there appears to be a larger change in $\Delta \sigma_1$ due to changes in interband transitions for $e\parallel B\parallel c$ than for $e\parallel B\parallel a$.   

As with $e\parallel B\parallel a$, the Drude weight increases more slowly at high fields, becoming sublinear above 5T. The oscillator strength associated with IBTs in W2 for $e\parallel c$ decreases at higher frequency, as shown by the decreasing slope in Fig. \ref{fig:conductivity}(b). This decreasing slope is an indication of the approach to the Lifshitz saddle point \cite{Jenkins2010}. Consequently, less IBT spectral weight is lost at higher fields, resulting in smaller increases to the Drude weight.

\section{Summary}
Chiral pumping is observed at THz frequencies in TaAs bulk single crystals in for fields oriented along both the $a$- and $c$- axes. Magneto-reflectance spectra of a TaAs bulk single crystal measured in the e$\parallel B\parallel a$ and $e\parallel B \parallel c$ Voigt geometries show blue-shifts of the screened plasma frequency with increasing magnetic field. These blues-shifts originate from Drude weight enhancements that are spectrally resolved from scattering rate effects. The enhancements agree with theoretical predictions of chiral pumping for TaAs. The Drude weight enhancement is accompanied by a reduction of the IBT spectral weight in accordance with the $f$-sum rule, providing further confirmation of Drude enhancement. The equivalence of this transfer by the $f$-sum rule was theoretically used to derive the same chiral pumping effect predicted by other theoretical methods. The spectral observation of this transfer therefore strongly validates the interpretation that chiral pumping is responsible for the field-dependence of the reflectance spectra.

At higher fields, the Drude weight increases more slowly with field, which is interpreted as a decreasing Fermi velocity along the field. These results are shown to follow naturally from the $f$ sum rule, and offer an explanation for the observed increase in longitudinal magneto-resistance at high fields in transport experiments.

\acknowledgements
  The authors would like to thank Jay D. Sau, Thomas E. Murphy, and Francois Joint for insightful discussions. Work at UCLA was supported by the U.S. Department of Energy (DOE) under Award Number DE-SC0011978. A. B. Sushkov and H. D. Drew were supported by DOE DE-SC0005436 grant, A. L. Levy, G.S. Jenkins and H. D. Drew were supported by NSF DMR-1610554. F. G. Liu was supported by China Scholarship Council No. 201508340005. 

\appendix
\section{Double Dirac Landau Level Spectrum and Selection Rules}

A double Weyl cone system is represented in the vicinity of the node by the following Hamiltonian:
\begin{equation}
\label{Hamiltonian}
\begin{aligned}
\hat{H}=\frac{\hbar^2 k_z^2}{2m^*}\hat{\sigma}_0+\left(-\frac{\Delta}{2}+\frac{\Delta}{2 k_0^2}k_z^2\right)\hat{\sigma}_z\\
+\left(2\Theta(k_z)-1\right) \left( \hbar v_x k_x \hat{\sigma}_x+\hbar v_y k_y \hat{\sigma}_y\right)
\end{aligned}
\end{equation}
where $\Delta$ is the Lifshitz gap, the Weyl points are located at $k=\left(0,0,±k_0\right)$, $v_x$ and $v_y$ are the Fermi velocities in the x- and y-directions, respectively, $\Theta(x)$ is a heavy-side function, and the operators $\hat{\sigma}_i$ are the Pauli spin matrices, with $\hat{\sigma}_0=I_{2\times2}$. Asymmetry between valence and conduction bands are introduced through massive term. 

\begin{figure}[htpb]
	\includegraphics[width=\columnwidth]{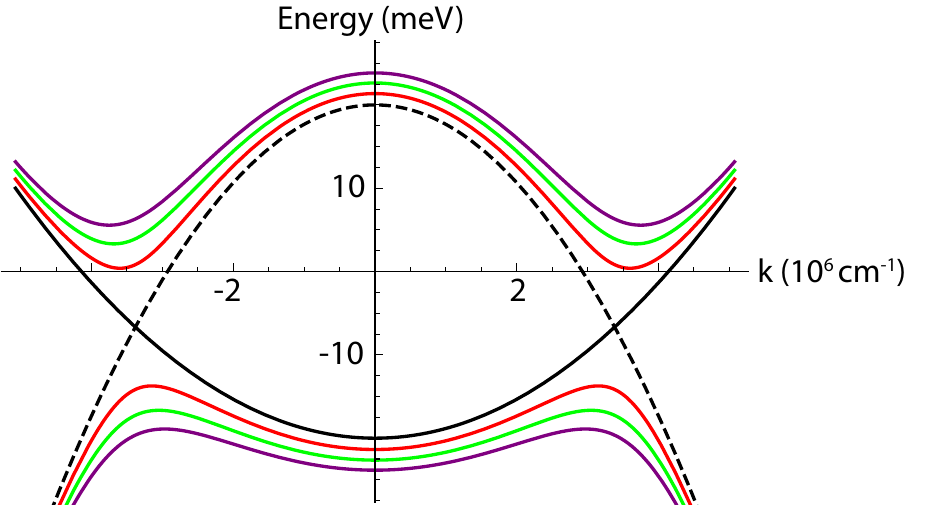}
	\caption{ (Color online). Shown is the dispersion of the lowest four LL's  predicted for a W2 Weyl pocket with magnetic field B=6T pointing along the line connecting the nodes. Equation \ref{Hamiltonian} is used  with parameters $\Delta=80$ meV,  $k_0=3.36\times 10^6$ cm$^{-1}$, $v_x=1.9\times10^7$ cm/s, $v_z=1.5\times 10^7$~ cm/s and $m^*=-0.3$ $m_e$ extracted from Ref. \cite{LeeBansil2015}. The Weyl point is located at $k_z=k_0$. The $n=0$ LL is depicted as either the dashed or solid black plots. The dipolar-like orientation of the chirality associated with the nodal pair determines which $n=0$ LL is a solution. } 
	\label{fig:Dispersion}
\end{figure}

\subsection{The Optical $f$-Sum Rule for Ideal Weyl Nodes}
As discussed in the main part of the paper, the $f$-sum rule demands that the total integrated spectral weight of our system remain constant with magnetic field. Losses in IBT spectral weight are offset by increases in the Drude weight.

For simplicity, the discussion will be limited to a Weyl node with perfectly linear dispersion in the x-, y-, and z-directions in the quantum limit. Combining Eqs. S4, and S7, the IBT spectral weight is:
\begin{widetext}
	\begin{equation}
	\label{eq:FSumProof}
	\begin{aligned}
	\int_0^{\Omega_{cut}}\sigma_1 (\omega)d\omega&=
	\int_0^{\Omega_{cut}}\sum_{n= 1}^{\frac{\Omega_{cut}^2 l_B^2}{8 v_x v_y}}\int_{-\infty}^{\infty}\frac{\hbar^2\Gamma}{2\pi l_B^2} \frac{n\frac{2\hbar v_x v_y}{l_B^2}}{2\left(n\frac{2\hbar v_x v_y}{l_B^2}+\hbar^2v_z^2k_z^2\right)^\frac{3}{2}}\frac{e^2 v_z^2}{\left(\hbar\omega-2\sqrt{n\frac{2\hbar v_x v_y}{l_B^2}+\hbar^2v_z^2k_z^2}\right)^2+\hbar^2\Gamma^2} \frac{dk_z}{2\pi}d\omega\\	
	&=\frac{e^2 v_z}{48\hbar\pi v_x v_y}\Omega_{cut}^2-\frac{e^2 v_z}{8\hbar\pi l_B^2} +\frac{e^2 v_z v_x v_y}{8\hbar\pi l_B^4 \Omega_{cut}^2}+ O\left(\frac{1}{\Omega_{cut}^4}\right)
	\end{aligned}
	\end{equation}
\end{widetext}
$\Omega_{cut}$ is a cut-off frequency much higher than the characteristic cyclotron energies of the Weyl bands. $\Omega_{cut}$ is large enough that terms of order $1/\Omega_{cut}^2$ are negligible. Since the Weyl node is assumed to be in the quantum limit, no IBTs are Pauli blocked. 

The first term of the last line in Eq. \ref{eq:FSumProof}, $\frac{e^2 v_z}{48\hbar\pi v_x v_y}\Omega_{cut}^2$, is equal to the integrated spectral weight of IBTs of this ideal Weyl node under zero magnetic field. The $f$-sum rule predicts the integrated spectral weight of the Drude to be $\frac{e^2 v_z}{8\hbar\pi l_B^2}$, equivalent to Eq. 2 in the main text.

The third term in Eq. \ref{eq:FSumProof}, $\frac{e^2 v_z v_x v_y}{8\hbar\pi l_B^4 \Omega_{cut}^2}$, relates to the IBT loss associated with the Drude enhancement expected in the semiclassical limit ($\mu l_B^2/8 v_x v_y>>1$) in ideal Weyl pockets \cite{Son2013,Hofmann2016}. This can be obtained by subtracting the spectral weight of IBTs at frequencies $\omega\leq 2\mu/\hbar$ from Eq. \ref{eq:FSumProof}.

Lastly, note that applying this derivation to the x- or y-directions yields analogous results. The fields do not need to be oriented along the z-axis in order to observe Drude enhancement associated with chiral pumping. 
\section{Faraday Geometry}

\begin{figure}[htpb]
	\includegraphics[width=3 in]{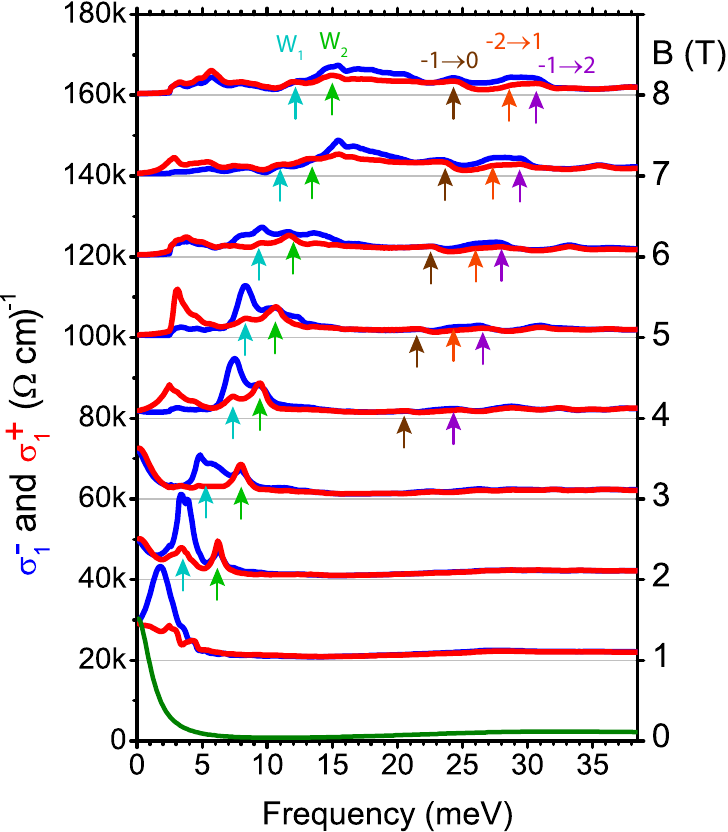}
	\caption{ (Color online). The optical conductivity, reported in the left (blue) and right (red) circular polarization basis, is derived from reflectance spectra measured in the $B\parallel c$ Faraday geometry. The peaks in conductivity denoted by cyan and green arrows are attributed to the onset of W1 and W2 cyclotron modes, respectively. The brown, orange, and purple arrows denote IBTs, where negative indices refer to LLs in the valence band.}
	\label{fig:Faraday}
\end{figure}

Reflectance spectra measured in the Faraday geometry with $B\parallel c$ are presented in Fig. \ref{fig:Faraday}. The scope of this section is confined to experimental results related to the quantum limit of W1 and W2. Uncertainty in the band parameters, asymmetry, and isotropy of all carrier pockets limit a complete analysis. Theoretical calculations of the optical selection rules (and values of the matrix elements to gauge expected strength of the transitions) involve these uncertainties in the bands. The anisotropy of identical Fermi pockets with multiple orientations in the Brillouin zone with respect to the applied magnetic field (parallel and perpendicular to the line connecting the nodes) presents further challenges in deciphering the LL transitions appearing in the data set. In short, the size, shape, polarization response, and number of the spectral resonances expected from multiple LL transitions in multiple pockets in diverse orientations are not well-known and make a complete analysis difficult. 

Many of these complications are avoided by limiting the scope of the analysis. The goal is to track the lowest-energy (cyclotron) transition as a function of field, and identify the optical signature expected when the W1 and W2 lowest Landau level transitions enter the quantum limit. The hallmark signature of a Weyl state entering the quantum limit is a spectrally broadened mode produced by transitions between the n=0 and the $n=\pm 1$ LLs.

Figure \ref{fig:Faraday} reports the optical conductivity extracted from relative reflectance spectra in the Faraday geometry with $B\parallel c$. The optical conductivity in the circular polarization gauge, $\sigma_1^+$ and $\sigma_1^-$, were obtained by taking the ratio of reflectance whose initial polarization (aligned along the $a$-axis) is rotated from the incident polarization by 45 degrees in each direction \cite{KuzmenkoRotation}. Since the circular polarization gauge is not a good basis for materials with anisotropic Fermi pockets, distinguishing holes from electrons in TaAs using the polarity of the cyclotron resonance in magnetic field in the circular polarization basis is not exact as the correct basis for a single ellipsoidal pocket is elliptically polarized light \cite{palik1970infrareda}. The absence of a strong difference between the $\sigma_1^+$ and $\sigma_1^-$ response in the W2 cyclotron ($0\rightarrow 1$) modes is attributed to anisotropy of the pockets in the $ab$-plane. 

\vspace*{\baselineskip}
The peaks in Fig.~ \ref{fig:Faraday}, indicated with cyan and green arrows, are attributed to cyclotron modes in the Weyl pockets. As a Weyl pocket enters the quantum limit, the optical conductivity spectrum is expected to broaden due to the large spectral range of allowable transitions between the $n=0$ and $n=\pm1$ LLs arising from the $k_z$-dependence of the Landau level spacing \cite{ashby2014PRB,shao2016magneto}. This is observed for the W2 cyclotron transition for $B\geq 2$ T, indicating a quantum limit in these pockets for $B< 2$ T. This finding is consistent with our conclusion that the W2 Fermi energy is below 3 meV as discussed in the paper and calculations of the Fermi velocity of the W2 pockets in the $ab$-plane \cite{LeeBansil2015,arnold2016}. Peaks attributed to W1 cyclotron modes are denoted by cyan arrows in Fig. \ref{fig:Faraday}. These peaks, present on the background of the W2 broadened peaks, do not appear to significantly broaden at fields below 7 T.

The peaks labeled with orange arrows are attributed to -2$\rightarrow$1 IBTs in the W1 pocket, as depicted in Figure \ref{fig:Faraday}. The -2$\rightarrow$1 transition is first observed at $B=5$ T, indicating that the bottom of the $n=1$ LL in the conduction band approaches the Fermi energy \cite{ashby2014PRB,shao2016magneto}. The -1$\rightarrow$2 IBT are denoted by purple arrows. The energy difference between -1$\rightarrow$2 and -2$\rightarrow$1 IBTs are attributed to asymmetry between valence and conduction bands of W1, as depicted in Fig. 4 of Ref. \cite{shao2016magneto}. Such asymmetry is consistent with DFT calculations \cite{LeeBansil2015,arnold2016}. The fact that the peak attributed to -2$\rightarrow$1 IBTs increases in intensity relative to that of the -1$\rightarrow$2 IBTs up to B=7T indicates that the quantum limit lies above $B = 6$ T \cite{shao2016magneto}, corroborating our conclusion from the magnetic field-dependence of the cyclotron modes. These findings are also consistent with our estimate that the Fermi energy is approximately 15 meV as discussed in the paper and values from band structure calculations \cite{LeeBansil2015,arnold2016}.
\section{Further discussion of zero-field broadband and Voigt fits}
\begin{table*}[htpb]
	\includegraphics[width=\textwidth]{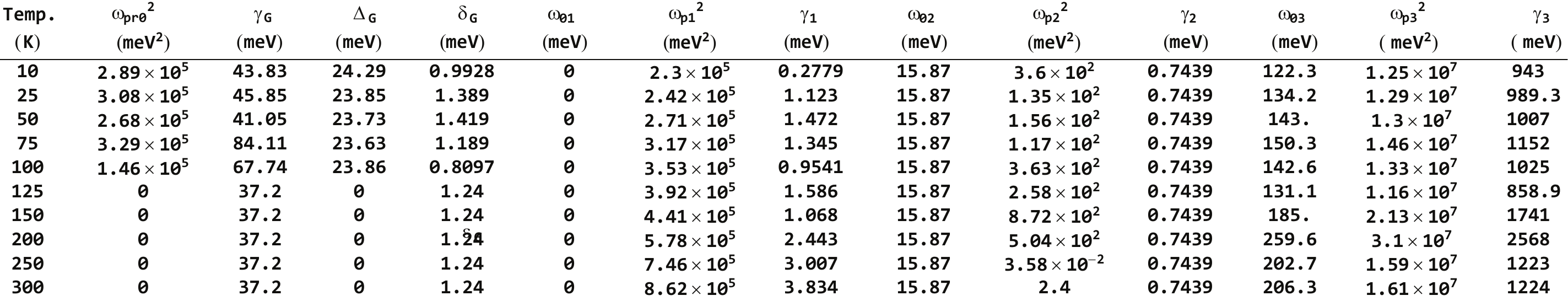}
	\caption{The fit parameters for zero field reflectance in the e$\parallel$a geometry are reported and the response functions plotted in Fig. \ref{fig:conductivity}(a). These parameters enter into the dielectric function through Eq. (1) of the main text.} 
	\label{aaxis0Field}
\end{table*} 
\begin{table*}[htpb]
	\includegraphics[width=\textwidth]{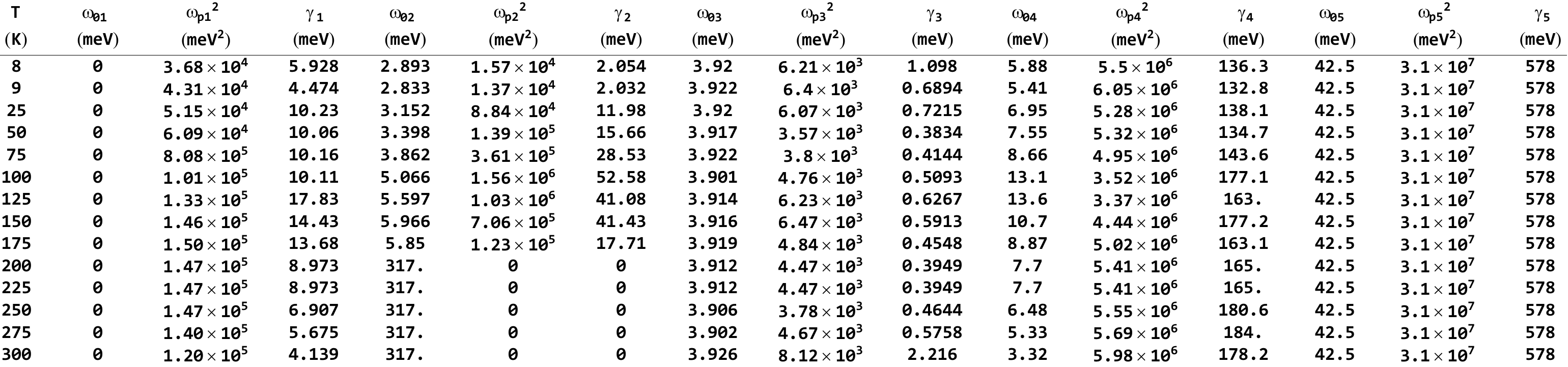}
	\caption{
		The fit parameters for zero field reflectance in the $e\parallel c$ geometry are reported and the response functions plotted in Fig. \ref{fig:conductivity}(b). These parameters enter into the dielectric function through Eq. (1) of the main text.} 
	\label{caxis0Field}
\end{table*}
\begin{table}[htb]
	\includegraphics[width=\columnwidth]{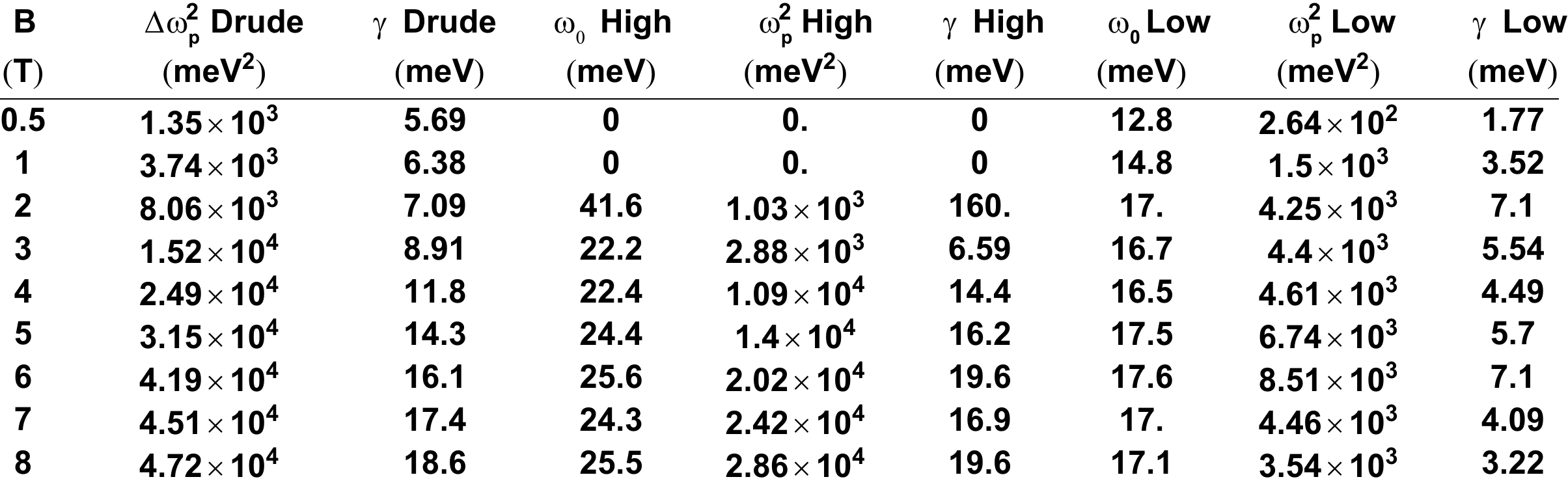}
	\caption{The fit parameters for the physical model for the $e\parallel B\parallel c$ optical response are reported. The zero-field $e\parallel c$ Drude dielectric term is replaced with a new Drude term with two free fit parameters ($\omega_p^2,\gamma_0$), and two negative Lorentzians each with three free parameters ($\omega_{0j},\omega_{sj}^2,\gamma_{j}$).} 
	\label{Drude+2LorentzianModel}
\end{table}
The parameters from optical fits (Tables \ref{aaxis0Field}-\ref{Drude+2LorentzianModel}) are reported in energy units for all fits with more than one degree of freedom. The scattering rate $\gamma$ in energy units is related to the transport lifetime $\tau$ as $\gamma=\hbar/\tau$.Tables \ref{aaxis0Field} and \ref{caxis0Field} refer to the parameters used to fit the zero-field spectra for $e\parallel a$ and $e\parallel c$, respectively. Table \ref{Drude+2LorentzianModel} shows the parameters for the dielectric model used to fit changes in the reflectivity spectra $\Delta R/R_0$ in the $e\parallel B\parallel c$ Voigt geometry for 0.5T $\leq B\leq$ 8T.  

In this model,the $e\parallel c$ zero field dielectric Drude term is replaced with a new Drude term with two free parameters ($\Delta (\omega_P^2)$ and $\gamma$), and adds two negative Lorentzians, each with three free parameters. The parameters are given in Table \ref{Drude+2LorentzianModel}. The fit to the reflectance data is shown in Figure 1(e) in the main text. The positive Drude and two negative oscillators are individually shown in Fig. \ref{fig:conductivity}(b) and the sum of the two oscillators are shown in Fig. \ref{fig:conductivity}(d). Spectral weight is removed from finite frequency oscillators, continuously increasing in magnitude and extending range to higher frequencies with field, and transferring to the Drude.

Representing the response as a single average scattering rate is a simplification as the Drude term may be more complicated due to the depth-dependent scattering from the surface roughness. The field dependence of the IBT optical response due to Landau level formation is more complex than the broad fit with two Lorentzian oscillators used in this work. 
However, the fits capture the main features of the $\Delta R/R_0$ data over the measured spectral range and are consistent with the change in optical conductivity obtained by the Kramers-Kronig method. The Drude and IBT oscillators demonstrate reasonable behavior. The Drude scattering rate and spectral weight increase with field, and the spectral weight of the IBTs lead to negative contributions to the conductivity that broaden and increase to higher frequencies with field.


\section{Obtaining $\Delta\varepsilon$ directly from $\Delta R/R_0$}

Because the real and imaginary parts of $\varepsilon(\omega)$ are subject to the Kramers-Kronig relation, the complex optical response from a reflectance spectrum can be obtained from:\cite{KKmethod}
\begin{equation}
\label{Reflection1}
\left(\frac{n(\omega)-1}{n(\omega)+1}\right)^2=R\textnormal{e}^{i\theta(\omega)}
\end{equation}
where $R$ is the reflectance, $n=\sqrt{\varepsilon(\omega)}$ is the complex index of refraction, and $\theta(\omega)=\frac{2\omega}{\pi}\int_0^\infty\frac {ln(R(\omega'))}{\omega'^2-\omega^2}d\omega'$.

The change in optical response from the relative change in reflectance $\Delta R(\omega)/R_0(\omega)$ can be obtained from the following equation:
\begin{equation}
\label{Reflection2}
\begin{array}{ll}
\left(\frac{n_0(\omega)+\Delta n(\omega)-1}{n_0(\omega)+\Delta n(\omega)+1}\right)^2&=(R_0(\omega)+\Delta R(\omega))\textnormal{e}^{i(\theta_0(\omega)  +\Delta \theta(\omega))}\\
\left(\frac{\frac{\Delta n(\omega)}{n_0(\omega)-1}+1}{\frac{\Delta n(\omega)}{n_0(\omega)+1}+1}\right)^2&=\left(1+ \frac{\Delta R(\omega)}{R_0(\omega)}\right)\textnormal{e}^{i\Delta \theta(\omega)}
\end{array}
\end{equation}
The change in $\theta(\omega)$ is given by:
\begin{equation}
\label{deltatheta}
\begin{array}{ll}
\Delta\theta(\omega)&=\frac{2\omega}{\pi}\int_0^\infty\frac {\textnormal{ln}\left(1+ \frac{\Delta R(\omega')}{R_0(\omega')}\right)}{\omega'^2-\omega^2}d\omega'
\end{array}
\end{equation}

The relative reflectance spectra shown in the main text are taken in the only spectral range in which $\Delta R/R_0$ is non-negligible. For $e \parallel B \parallel a$, $\Delta R/R_0$, uncertainty in $\Delta\varepsilon(\omega)$ in the measured spectral range form neglecting contributions from $\Delta R/R_0$ outside of the spectral range are negligibly small. For $e \parallel B \parallel c$, small uncertainty in the change in optical response is introduced due to $\Delta R/R_0$ at frequencies below the measured spectral range.

\bibliography{TI}

\end{document}